\documentstyle[twocolumn,prl,aps]{revtex}
\input{epsf.tex}
\def\inseps#1#2{\def\epsfsize##1##2{#2##1} \centerline{\epsfbox{#1}}}

\begin{document}
\def \beq{\begin{equation}}
\def \eeq{\end{equation}}
\def \beqarr{\begin{eqnarray}}
\def \eeqarr{\end{eqnarray}}

\twocolumn[\hsize\textwidth\columnwidth\hsize\csname @twocolumnfalse\endcsname

\draft

\title{
Evidence of Charge Density Wave Ordering in Half Filled High Landau
Levels
}

\author{E. H. Rezayi$^a$, F. D. M. Haldane$^b$, and Kun Yang$^c$}
\address{
$^a$Department of Physics, California State University, Los Angeles, 
California 90032}

\address{
$^b$Department of Physics, Princeton University, Princeton, New Jersey 08544}

\address{
$^c$Condensed Matter Physics 114-36, California Institute of Technology,
Pasadena, California 91125}

\date{\today}

\maketitle
\begin{abstract}
We report on numerical studies of two-dimensional electron systems 
in the presence of a perpendicular magnetic field, 
with a high Landau level (index $N\ge 2$) half filled by 
electrons. 
Strong and sharp peaks are found in the wave vector dependence of
both static density susceptibility and equal-time
density-density correlation function, in finite-size
systems with up to twelve electrons. 
Qualitatively different from partially filled lowest ($N=0$) Landau level,
these results are
suggestive of a tendency toward charge
density wave ordering in these systems. The ordering wave vector 
is found to decrease with increasing $N$. 

\end{abstract}

]

Two-dimensional (2D) electron gas systems subject to a perpendicular
magnetic field display remarkable phenomena, reflecting the importance
of electronic correlations. The most important
among them is the fractional quantum Hall effect (FQHE), which was found
in the strong field limit, where the electrons are confined 
to the lowest ($N=0$) or the second ($N=1$) Landau levels. The physics of
FQHE is reasonably well understood\cite{books}: the kinetic energy
of the electrons are quenched by the strong perpendicular
magnetic field and the Coulomb interaction dominates the physics of the
partially filled Landau level; 
at certain Landau level filling factors ($\nu$,
defined to be the ratio of 
the number of electrons
to the number of Landau orbitals in each Landau level) 
the electrons condense into a highly-correlated, incompressible quantum
fluid, giving rise to quantized Hall resistivity ($\rho_{xy}$) and 
thermally activated longitudinal resistivity ($\rho_{xx}$).

Experimentally, 
the FQHE has never been found at filling factors $\nu > 4$, when the partially
filled Landau level has Landau level index $N \ge 2$ (taking into account the
two spin species of the electrons). Nevertheless, recent 
experiments\cite{lilly,du}
on high quality samples have revealed remarkable transport
anomalies for $\nu > 4$, especially when $\nu$ is near a half integer, which
means the partially filled Landau level is nearly half filled. Such 
anomalies include a strong anisotropy and nonlinearity in 
$\rho_{xx}$. They reflect intriguing correlation physics at work in these
systems that is qualitatively different from  
the FQHE and yet to be completely understood.

It was argued\cite{anderson}, before the discovery of the FQHE, that
the ground state of a 2D electron gas in a strong magnetic field 
may possess charge density wave (CDW) order.
Recent Hartree-Fock (HF) calculations\cite{koulokov,fogler,moessner}
find that single-Slater
determinant states with CDW order have energies lower
than the Laughlin-type liquid states for $N\ge 2$. The CDW state with 1D
stripe order, or stripe phase\cite{koulokov,fogler,moessner},
which is predicted to be stable near half filling for the partially filled
Landau level, can in principle give rise to transport anisotropy as the
orientation of the stripe picks out a special direction in space.
Questions remain, however, with regard to the stability of the HF states 
against quantum fluctuations as well as disorder, especially when $N$ is not
too large.

In this paper we report on results of numerical studies of
partially filled Landau levels with $N \ge 2$, in finite size systems with
up to $N_e=12$ electrons and torus geometry. The torus is ideal for this study,
because, while it maintains translational and rotational invariance in the plane
for the
infinite system, it does break them for finite sizes and therefore  
produces a preferred direction that allows for the CDW state to be aligned 
in one direction. The spherical geometry (which is another popular geometry for
finite size studies), on the other hand, will rotationally average 
the state and necessarily introduce defects.
We assume the magnetic field is
sufficiently strong so that the filled Landau levels are completely inert 
and mixing between different Landau levels can be neglected. 
We calculate numerically the energy spectra,
the wave vector dependence of the 
static density susceptibility ($\chi(q)$) and the 
density-density correlation in 
the ground state ($S_0(q)$). 
We find strong and sharp peaks in both $\chi(q)$ and $S_0(q)$.
These results are strongly suggestive of a tendency toward charge
density wave ordering in the ground state.
In accordance with this, we find nearly degenerate low-energy states that
are separated by the ordering wave vector.

The methods used in the present study are identical to those used in numerical
studies of the partially filled lowest Landau level\cite{haldane}.
Since the kinetic energy is quenched by the magnetic field, the Hamiltonian
contains the Coulomb interaction alone, which,
after projecting onto the $N$th Landau 
level\cite{haldane}, takes the form 
\beq
H=\sum_{i<j}\sum_{\bf q}e^{-q^2/2}[L_N(q^2/2)]^2V(q)e^{i{\bf q}\cdot(
{\bf R}_i-{\bf R}_j)},
\eeq
where ${\bf R}_i$ is the guiding center\cite{haldane}
coordinate of the $i$th electron, $L_N(x)$ are the 
Laguerre polynomials, and $V(q)=2\pi e^2/q$ is the Fourier transform of the
Coulomb interaction. The magnetic length $\ell$ is set to be 1 for convenience.
In this work we impose periodic boundary conditions in both $x$ and $y$ 
directions
in the finite-size system under study, and the ${\bf q}$'s are wave vectors
that are compatible with the size and geometry of the system.
We diagonalize the above Hamiltonian exactly 
and calculate various correlation functions.

\begin{figure}
\inseps{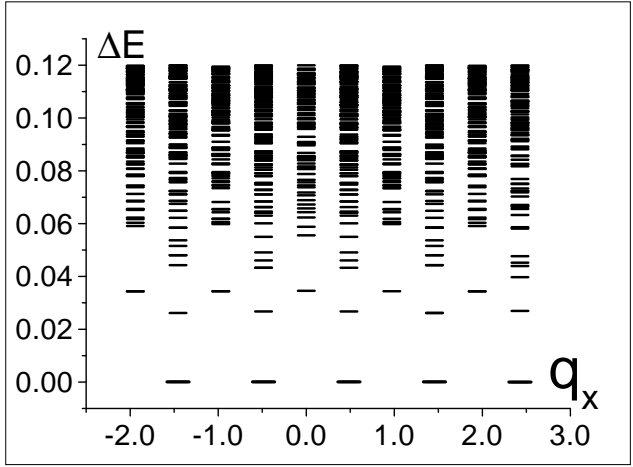}{0.34}
\caption{Energy (in unit of $e^2/\epsilon\ell$)
versus the x-component of the momentum 
for the half-filled $N=2$ Landau level with 
ten electrons, rectangular geometry, and aspect ratio 0.75. The
momenta of the five nearly degenerate low-energy states are
$(\pm 0.485, 0), (\pm 1.456, 0)$, and $(2.427, 0)$.}
\end{figure}

In Fig. 1 we present the energy spectrum of 
half-filled $N=2$ Landau level with
ten electrons, rectangular geometry, and aspect ratio 0.75.
The spectrum is qualitatively different from the known incompressible
FQHE states in the following aspects: i) the ground state is not separated
from the excited states by a gap; instead, there are several nearly degenerate
low energy states, separated by a characteristic wave vector ${\bf q}^*$ (the
physical importance of ${\bf q}^*$ will be discussed later). 
This almost exact degeneracy is not specific to this particular geometry; 
in Fig. 2 we have plotted the energy levels (for all momenta) versus aspect
ratio $a$, in which it is clear that the near degeneracy is present for
$0.7 < a < 1.0$.
A gap
in the spectrum, on the other hand,
is the most important property of a FQHE state that gives rise to
incompressibility. ii) The momentum of the ground state is, in general, 
not related to any reciprocal lattice vectors and is sensitive to 
the geometry of the system; such sensitivity to boundary conditions is 
characteristic of compressible states. The known FQHE states, however, all
have ground state momentum equal to one half of a reciprocal lattice vector and
 are independent of system geometry, reflecting intrinsic topological
properties of the state.
The spectra and quantum numbers of these ground states are also very different
from
the Fermi-liquid like compressible state at half filling in the lowest
Landau level\cite{hlr}. In that case 
cluster-like ansatz
variational wave functions\cite{rr,haldane1} have remarkably large 
overlaps with the exact ground state and, from the shape of the
optimal cluster for a given geometry, 
{\em predict} the momentum of the ground state with remarkable
accuracy\cite{haldane1}. The momenta of the ground states here, however,
do not match those of the optimal cluster states. Important differences
are also seen in the response functions studied below. We thus conclude that
these states are compressible and have properties that are qualitatively
different from the Fermi-liquid like compressible state at half filling in the
lowest Landau level.

\begin{figure}
\inseps{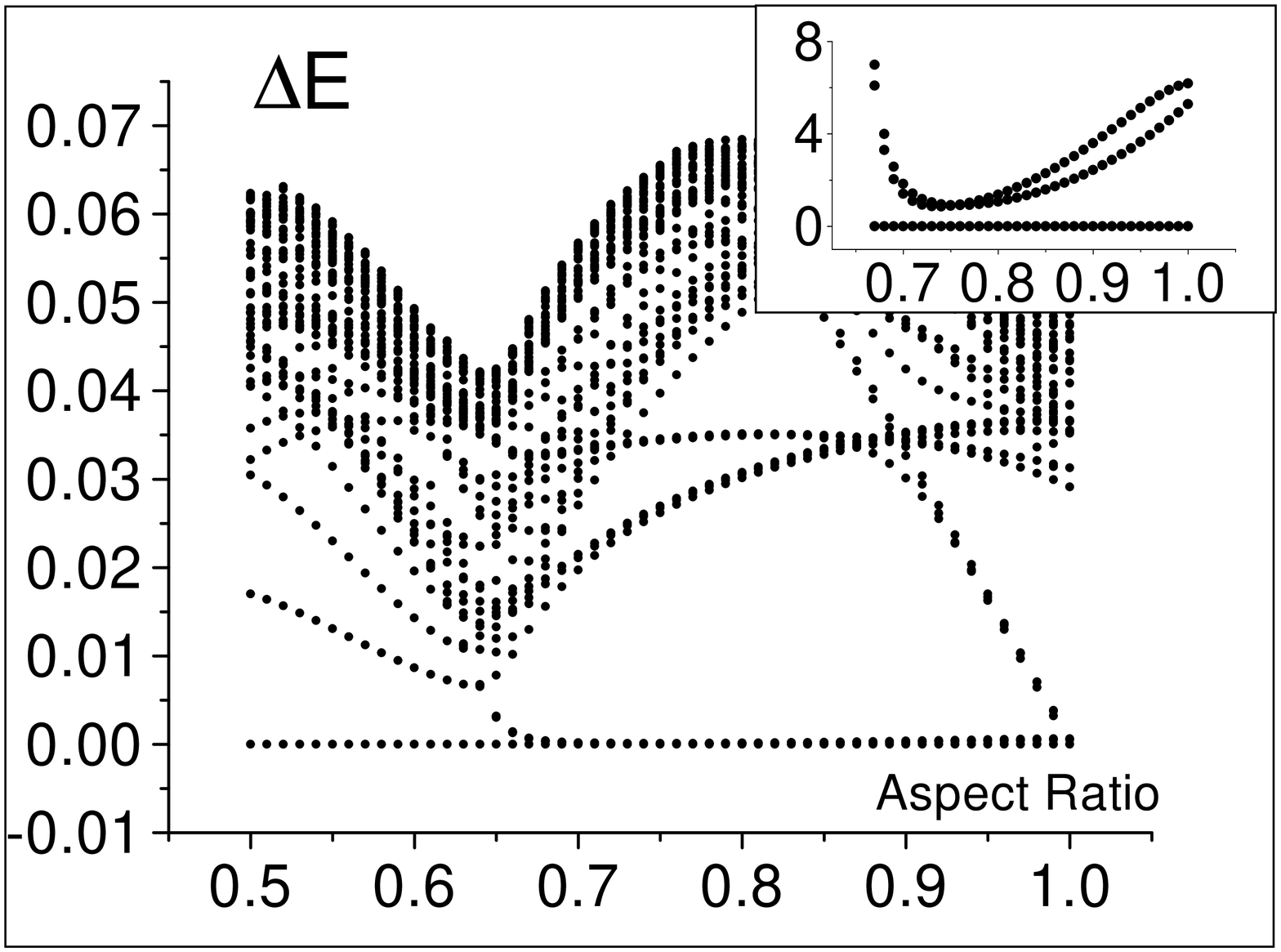}{0.36}
\caption{Energy levels versus aspect ratio for half-filled $N=2$ Landau level
with ten electrons and rectangular geometry. The inset is a blow-up 
of the low-energy spectra for aspect ratio between 0.7 and 1.0; the energies
have been multiplied by $10^4$.}
\end{figure}

We now turn to the discussion of response functions.
Here the fundamental quantity of interest is the dynamical structure factor,
defined to be
\beq
S_0({\bf q}, \omega)
={1\over N_e}\sum_n|\langle 0|\sum_ie^{i{\bf q}\cdot{\bf R}_i}
|n\rangle|^2\delta(E_n-E_0-\omega).
\eeq
The summation is over states in the given Landau level that is being
studied.
Various physically important quantities may be calculated from 
$S_0({\bf q}, \omega)$\cite{haldane};
in particular, the projected static density response function is the inverse 
moment of $S_0({\bf q}, \omega)$:
\beq
\chi({\bf q})=\int_0^{\infty}d\omega S_0({\bf q}, \omega)/\omega,
\eeq
and the projected equal time density-density correlation function is the
$0$th moment of $S_0({\bf q}, \omega)$:
\beq
S_0({\bf q})=\int_0^{\infty}d\omega S_0({\bf q}, \omega).
\eeq

\begin{figure}
\inseps{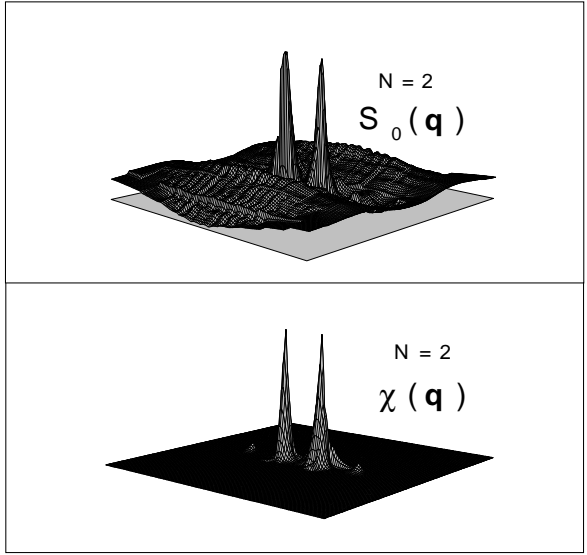}{0.36}
\caption{$S_0({\bf q})$ and $\chi({\bf q})$ (in unit of ${\epsilon\ell/e^2}$)
for half-filled $N=2$ Landau level with ten electrons. The data were taken from
systems with rectangular geometry, and a number of different aspect ratios.
The highest peak value for $S_0({\bf q})$ and $\chi({\bf q})$ 
are 3.7 and 39726.3 respectively, which occur at aspect ratio 0.75.
}
\end{figure}

In Figs. 3 and 4 we present the ${\bf q}$ dependence of $\chi({\bf q})$ and
$S_0({\bf q})$ for half filled $N=2$ and $N=3$ Landau levels
respectively. The data were
taken from systems with 10 electrons
and rectangular geometry, for several different aspect ratios.
We see sharp and strong peaks in $\chi({\bf q})$ 
at ${\bf q}^*=(0.97, 0)$ and ${\bf q}^*=(0.84, 0)$ for 
$N=2$ and $N=3$ Landau levels respectively. At the
scale of the peak value of $\chi$, the response at different ${\bf q}$'s are
indistinguishable from zero, except for a secondary peak positioned at 
{\em exactly}
$3{\bf q}^*$, the height of which is much smaller but quite noticeable. 

This is strongly suggestive of a tendency toward CDW ordering, as an 
(unpinned) CDW responds strongly to an external potential
modulation with a wave vector that matches one of its reciprocal lattice 
vectors.
The fact that secondary peaks appear only at integer multiples of
the primary wave vector (${\bf q}^*$) suggests that the CDW has 1D (stripe) 
structure. The absence of response at $2{\bf q}^*$ 
(and any other even multiples)
is consistent with
the presence of particle-hole (PH) symmetry in the underlying Hamiltonian
at half filling: the PH transformation of the CDW state is equivalent to 
translation by half a period.  
Consistent with the stripe CDW picture,
sharp peaks are also observed
in $S_0({\bf q})$ at ${\bf q}={\bf q}^*$, 
indicating strong density-density correlation in
the ground state at the ordering wave vector. Similar behavior is also seen
at higher Landau levels and other filling factors.
Again, these results are in sharp contrast with FQHE states (like $\nu=1/3$
in the lowest Landau level), or the Fermi liquid-like state of the half
filled lowest Landau level. In the former case, $\chi$ is small 
(below 20) for all
$q$'s due to the large gap in the spectrum\cite{haldane}; 
while in the latter case although there are peaks in both $\chi(q)$ and 
$S_0(q)$, the peak position is tied to $q=2k_F=2$ and the height of 
the peak in $\chi(q)$ is in general much smaller than those seen here and
the peaks are broader\cite{haldane1}. 
 
\begin{figure}
\inseps{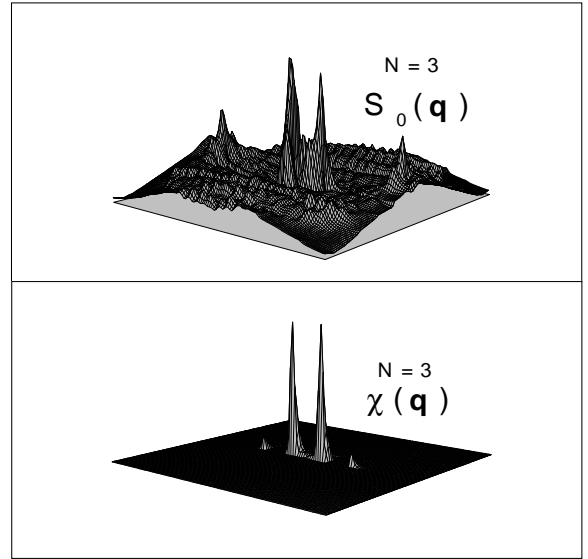}{0.36}
\caption{Same as in Fig. 3, for $N=3$ Landau level.
The highest peak values are 4.17 and 2953432.4 respectively, which occur at
aspect ratio 0.56.}
\end{figure}

The origin of the strong peaks in $\chi$ at ${\bf q}^*$ may be traced
to the almost exact 
degeneracy of the low energy states in the spectra that are separated
by ${\bf q}^*$. 
Since 
these states are connected by a potential modulation with wave vector
${\bf q}^*$, an extremely small energy denominator ensures huge response
from the system. We emphasize that such near degeneracy is a generic 
property of the system and not specific to a particular system size or 
geometry\cite{finitesize}.
This can be seen clearly in Fig. 1b, where we plot energy levels from 
different geometries together. If the system develops long-range CDW
order in the thermodynamic limit, we would
expect the number of nearly degenerate 
states to increase with system size; in the thermodynamic limit, there will
be infinitely many of them, spaced in momentum by ${\bf q}^*$, 
that become {\em exactly} degenerate.
In this case
a ground state that {\em spontaneously} breaks the translational symmetry
may be constructed by taking linear combinations
of these degenerate states, even though this is not possible (unless 
degeneracy is exact) in any finite
system where all eigenstates must have a definite momentum.

In Fig. 5 we plot the Landau level dependence of $q^*$ for half 
filled Landau levels. For systems with a 
given number of electrons in a given Landau level, we define $q^*$ to be
the modulus of the
wave vector that gives the largest $\chi({\bf q})$ for all geometries 
(aspect ratios).
Despite the small and non-systematic dependence on the number of electrons,
it is clear that $q^*$ decreases with increasing $N$. We are 
unable to accurately determine $q^*$ beyond $N=4$, as in this case $1/q^*$
becomes comparable to (and eventually exceeds) the linear size of the largest
size system that we are able to study. This result is qualitatively 
consistent with the prediction of Hartree-Fock 
theory\cite{koulokov,fogler,moessner}, which predicts that the period of the
CDW is set by the scale of the 
cyclotron radius in a given Landau level.
In units of inverse magnetic length,
this implies $q^*\propto 1/\sqrt{2N+1}$. 

\begin{figure}
\inseps{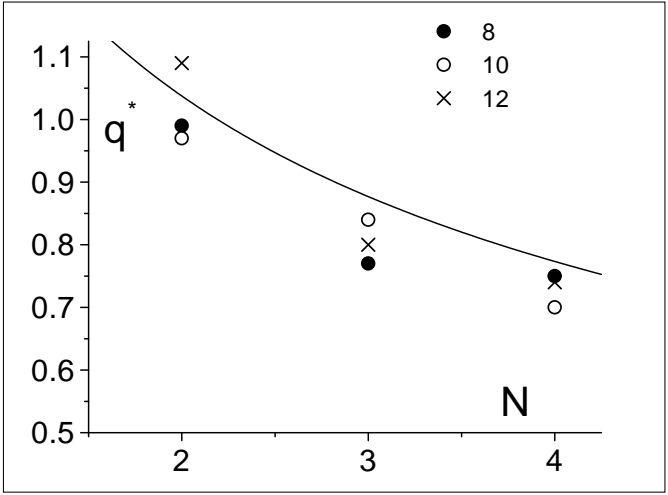}{0.36}
\caption{
Estimated ordering wave vector $q^*$ versus Landau level index $N$.
The solid curve is the prediction of Hartree-Fock theory.
}
\end{figure}

We have also performed numerical studies away from half-filling, 
at filling factors $\nu=1/4, 1/3, 2/5$, etc, for Landau levels
$2\le N \le 6$. At all these filling factors,
we have seen qualitatively similar behavior in $\chi$ and $S_0$, which
are suggestive of a tendency toward CDW ordering. No evidence of incompressible
FQHE states is found. This is consistent with Hartree-Fock theory, which
predicts\cite{fogler2} no FQHE for $N \ge 2$. Hartree-Fock theory also 
predicts that 
when sufficiently far away
from half-filling, the stripe phase becomes unstable
against a different type of CDW ordering, the ``bubble phase"\cite{fogler}, 
which has a 
two-dimensional lattice structure. In our calculations we have seen some
indication that this may be the case; however more work is needed to make a
clear distinction between these two types of structures. We leave this to 
future investigation.

As stated above, the physical properties of the systems that we have studied 
here are qualitatively consistent with predictions of Hartree-Fock theory.
We also find that in the single Slater determinant basis for the
wave functions, the Hartree-Fock single Slater determinant (with simple
stripe structure) has the highest
weight in the exact ground state.
For example, in $N=2$ and $N=3$ Landau levels with ten electrons and 
rectangular aspect ratios of 0.75 and 0.56 respectively, the 
highest weight single Slater determinant has the occupation pattern (in 
Landau gauge) $11111000001111100000$ (and ones that may be obtained by
translating this pattern), where $1$ stands for an occupied 
orbital and $0$ stands for an empty orbital. The maximum weights
are 0.1463 and 0.1986 for $N=2$ and $N=3$ Landau levels; the next highest weight
single Slater determinant has a weight that is approximately fifteen times 
smaller for $N=2$ and 1400 times smaller for $N=3$.
However, by making a linear combination of the five 
nearly degenerate  states we can
single out the above occupation pattern and construct an approximate  
eigenstate that breaks translational
symmetry.  The overlaps (squared) with HF wave function then
become 0.7338 for $N=2$ and 0.9930 for $N=3$.
While there are still some fluctuations
on top of HF states for $N=2$, these fluctuations are completely gone for
$N=3$.
Fradkin and Kivelson\cite{fradkin} recently considered the effects 
of thermal and quantum fluctuations on the Hartree-Fock stripe phase, 
and predicted a number of novel phases. Due to the 
limited system sizes in numerical studies, we are unable to
distinguish among these phases (whose distinctions show up at large distances
only).

We have benefited from discussions with Jim Eisenstein, 
Steve Girvin, and Mike Lilly. We thank Boris Shklovskii for 
helpful comments and for reminding us of
Ref. \cite{fogler2}.  This work was started at 
ITP Santa Barbara during
the {\em Disorder and Interactions in Quantum Hall and
Mesoscopic Systems} workshop, which was supported by NSF PHY-9407194. 
EHR was supported by NSF DMR-9420560, FDMH by NSF DMR-9809483,
and KY by the Sherman Fairchild foundation.


\begin{references}
\bibitem{finitesize} There are finite size effects, however, coming from the
fact that the system needs to adjust the periodicity of the CDW to be 
commensurate with the system size. Such effects become more pronounced below
$a=0.67$.
Detailed discussions of such effects will
be presented elsewhere.
\end{references}
\end{document}